\documentclass{article}
  \usepackage{emulateapj}
\input{psfig}

\lefthead{Brandner et al.}
\righthead{HST and VLT observations of PROPLYDS in the giant HII region NGC 3603}

\newcommand\NII{N{\sc ii}}

\begin{document}

\title{HST/WFPC2 and VLT/ISAAC observations of PROPLYDS in the giant HII region NGC 3603\footnote{Based on 
observations obtained at the European Southern Observatory, Paranal and La Silla
(ESO Proposal No.\ 47.5-0011, 53.7-0122, 58.E-0965, 59.D-0330, 63.I-0015), 
and on observations made with the 
NASA/ESA Hubble Space Telescope, obtained from the Space 
Telescope Science Institute. STScI is operated by the Association of 
Universities for Research in Astronomy, Inc., under the NASA contract NAS
       5-26555. }}

\author{Wolfgang Brandner\altaffilmark{1}, Eva K.\ Grebel\altaffilmark{2,3}
You-Hua Chu\altaffilmark{4}, Horacio Dottori\altaffilmark{5},
Bernhard Brandl\altaffilmark{6},
Sabine Richling\altaffilmark{7,8}, Harold W.\ Yorke\altaffilmark{7},
Sean D.\ Points\altaffilmark{4}, Hans Zinnecker\altaffilmark{9}}

\affil{$^1$University of Hawaii, Institute for Astronomy, 2680 Woodlawn Dr.,
Honolulu, HI 96822, USA}
\authoremail{brandner@ifa.hawaii.edu}

\affil{$^2$University of Washington at Seattle, Astronomy Department,
Box 351580, Seattle, WA 98195, USA}
\authoremail{grebel@astro.washington.edu}

\affil{$^3$ Hubble Fellow}

\affil{$^4$University of Illinois at Urbana-Champaign, Department of Astronomy,
1002 West Green Street, Urbana, IL 61801, USA}
\authoremail{chu@astro.uiuc.edu, points@astro.uiuc.edu}

\affil{$^5$Instituto de F\'{\i}sica, UFRGS, Campos do Vale, C.P. 15051, 
91500 Porto Alegre, R.S., Brazil}
\authoremail{dottori@if.ufrgs.br}

\affil{$^6$Cornell University, Department of Astronomy,
222 Space Sciences Building, Ithaca, NY 14853, USA}
\authoremail{brandl@astrosun.tn.cornell.edu}

\affil{$^7$Jet Propulsion Laboratory, California Institute of Technology, 
4800 Oak Grove Drive, Mail Stop 169-506, Pasadena, CA 91109, USA}
\authoremail{richling@ita.uni-heidelberg.de, Harold.Yorke@jpl.nasa.gov}

\affil{$^8$Institut f\"ur Theoretische Astrophysik, Universit\"at Heidelberg,
Tiergartenstra{\ss}e 15, D-69121 Heidelberg, Germany}

\affil{$^9$Astrophysikalisches Institut Potsdam, An der Sternwarte 16,
D-14482 Potsdam, Germany}

\begin{abstract} 
We report the discovery of three proplyd-like structures in the giant HII
region NGC 3603. The emission nebulae are clearly resolved in narrow-band
and broad-band HST/WFPC2 observations in the optical and broad-band
VLT/ISAAC observations in the near-infrared. All three nebulae are tadpole 
shaped, with the bright ionization front at the head facing the central 
cluster and a fainter ionization front around the tail pointing away 
from the cluster. Typical sizes are 6,000 A.U. $\times$ 
20,000 A.U. The nebulae share the overall morphology of the proplyds 
(``PROto PLanetarY DiskS'') in Orion, but are 20 to 30 times larger in size.
Additional faint filaments located between the nebulae and the
central ionizing cluster can be interpreted as bow shocks resulting
from the interaction of the fast winds from the high-mass stars in the cluster
with the evaporation flow from the proplyds.

Low-resolution spectra of the brightest nebula,
which is at a projected separation of 1.3 pc from the cluster, reveal
that it has the spectral excitation characteristics of an Ultra Compact 
HII region with electron densities well in excess of 10$^4$\,cm$^{-3}$.
The near-infrared data reveal a point-source superimposed on the
ionization front.
	
The striking similarity of the tadpole shaped emission nebulae in NGC 3603
to the proplyds in Orion suggests that the physical structure of both types
of objects might be the same. We present 2D radiation hydrodynamical
simulations of an externally illuminated star-disk-envelope system,
which was still in its main accretion phase when first
exposed to ionizing radiation from the central cluster.
The simulations reproduce the overall morphology  of the proplyds
in NGC 3603 very well, but also indicate that 
mass-loss rates of up to 10$^{-5}$ M$_\odot$\,yr$^{-1}$ are required in 
order to explain the size of the proplyds.

Due to these high mass-loss rates, the proplyds in NGC 3603 should only
survive $\approx$10$^5$ yr. Despite this short survival time,
we detect three proplyds. This indicates that circumstellar disks
must be common around young stars in NGC 3603 and that these particular
proplyds have only recently been exposed to their present harsh UV
environment.

\end{abstract}


\section{Introduction}

HST/WFPC2 observations of the Orion Nebula (M42) revealed a large
variety of dark silhouette disks (O'Dell \& Wong 1996;
McCaughrean \& O'Dell 1996) and
partially ionized circumstellar clouds (O'Dell et al.\ 1993). Many of the
circumstellar clouds, which had first been detected from the ground by 
Laques \& Vidal (1979), have a cometary shape with the tails
pointing away from the O7V star $\Theta^1$\,Ori C and the O9.5V star
$\Theta^2$\,Ori A, the
brightest and most massive members of the Trapezium cluster.
The partially ionized circumstellar clouds with cometary shape were
identified as protoplanetary disks (``proplyds'') around young stars, 
which are ionized from the outside (Churchwell et al.\ 1987;
O'Dell et al.\ 1993).

Many proplyds are ionization bounded, which indicates
that all EUV photons (h$\nu\ge$ 13\,eV) get absorbed in the ionization front
engulfing the protostar and its circumstellar disk
(O'Dell 1998).  FUV photons (13\,eV $>$ h$\nu$ $\ge$ 6\,eV), however,
are able to penetrate the ionization front. 
They heat up the inside of the proplyd envelope and lead to the
dissociation of molecules in the outer layers of the circumstellar disk 
(Johnstone et al.\ 1998). The resulting evaporation flow provides a steady 
supply of neutral atoms to the ionization front and leads to the
development of a cometary tail (McCullough et al.\ 1995; St\"orzer \& 
Hollenbach 1999). 

Because of their larger size and the ionized envelope, proplyds can
be spotted more easily than circumstellar disks themselves. 
Consequently, Stecklum et al.\ (1998) proposed to utilize proplyds as tracers
for circumstellar disks in distant star forming regions.
Systematic search efforts for proplyds in
HII regions around young clusters with WFPC2 did not yield any
new detections (Stapelfeldt et al.\ 1997; Bally et al.\ 1998a). 
Until recently, only
one other proplyd had been found. It is located in the vicinity of the 
O7V star Herschel 36 in the Lagoon Nebula (M16, Stecklum et al.\ 1998).

NGC 3603 is located in the Carina spiral arm at a
distance of 6\,kpc (De Pree et al.\ 1999 and references therein). 
With a bolometric luminosity L$_{\rm bol}>10^7$ L$_\odot$,
NGC 3603 is 100 times more luminous than the Orion Nebula and has about 10\%
of the luminosity of 30 Doradus in the Large Magellanic Cloud (LMC). It is 
the only Galactic giant HII region
whose massive central ionizing cluster can also be studied at optical 
wavelengths. The initial
mass function of the cluster follows a Salpeter type power law with index
$\Gamma$=$-$1.70 for masses $>$ 25\,M$_\odot$ and $\Gamma$=$-$0.73
for masses less than 25\,M$_\odot$ (Eisenhauer et al.\ 1998), extending
from Wolf-Rayet stars and O3V stars with masses up to 120\,M$_\odot$ 
(Drissen et al.\ 1995) down to stars of at least 1\,M$_\odot$ (Eisenhauer et 
al.\ 1998). The total cluster mass is $\ge$ 4,000\,M$_\odot$. 

To the south of the cluster is a giant molecular cloud.  Ionizing radiation
and fast stellar winds from the starburst cluster are excavating large
gaseous pillars.  Located about 20$''$ to the north of the cluster
center is the blue supergiant Sher 25. This supergiant is unique because
its circumstellar ring and bipolar outflows form an hourglass structure
similar to that of SN1987A (Brandner et al.\ 1997a, 1997b).

As part of a follow-up study on the hourglass structure around Sher 25 
we observed NGC 3603 with HST/WFPC2.
In this paper, we report the serendipitous discovery of three proplyd-like
structures in NGC 3603 based on HST/WFPC2 and VLT/ISAAC observation 
and perform a first analysis of their physical properties.

\begin{figure*}[ht]
\centerline{ \psfig{figure=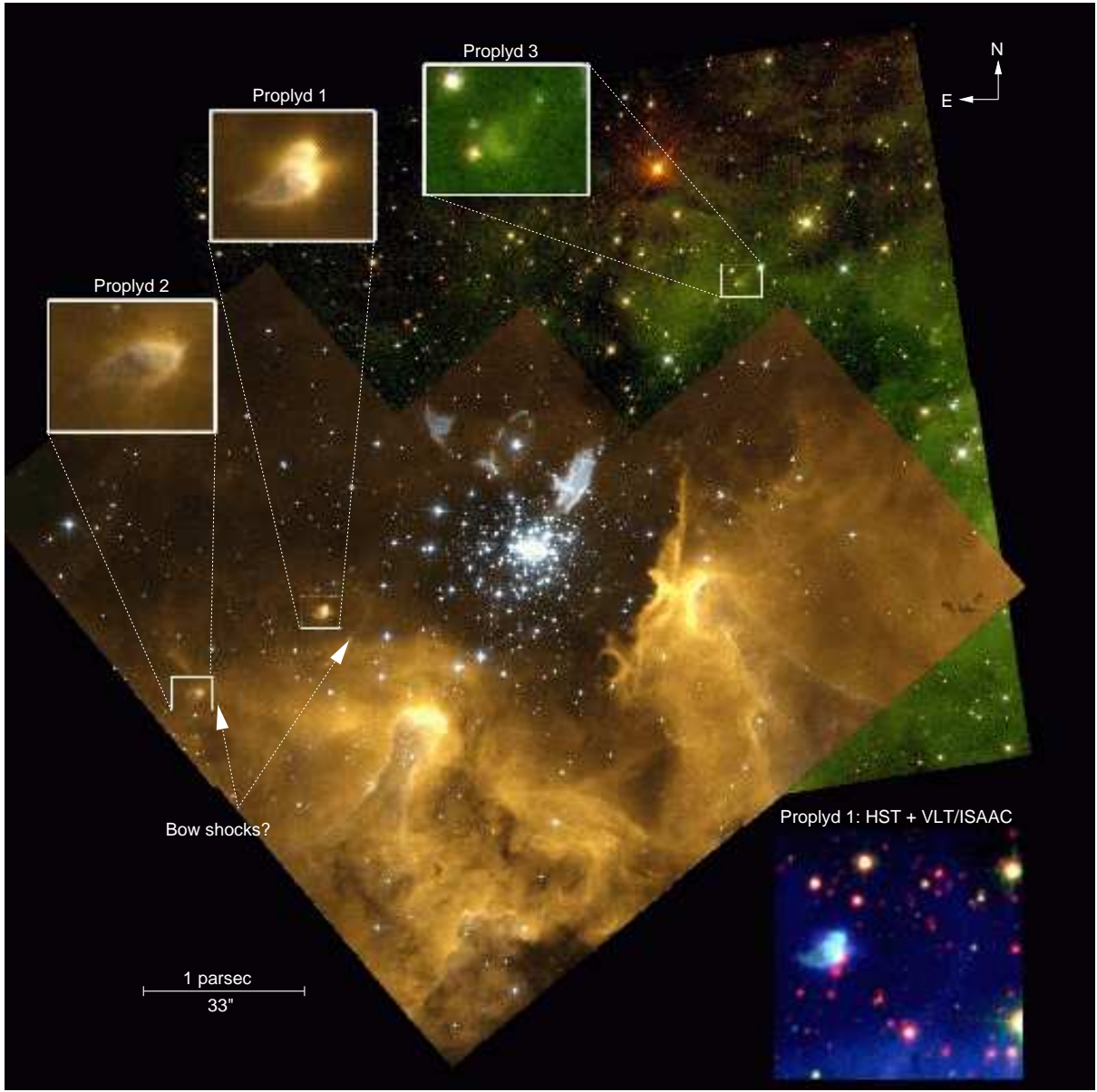,width=16cm}}
\caption{WFPC2 observations of NGC 3603. North is up and
east is to the left.
The upper part of the image consists of the archive data with the following
color coding: F547M (blue), F675W (green), F814W (red). Overlaid
are our new WFPC2 data  with the F656N data in the red channel,
the average of F656N and F658N in the green channel, and F658N
in the blue channel.
 The location of the
three proplyd-like emission nebulae is indicated.
The insert at the lower right is a combination of WFPC2 F656N (blue)
and F658N (green) and VLT/ISAAC K$_{\rm s}$ (red) observations.\label{fig1}}
\end{figure*}

\section{Observations and data reduction} 
\subsection{HST/WFPC2 observations}
On March 5, 1999 we obtained deep narrow-band
H$\alpha$ (F656N, 2$\times$500s) and [\NII] (F658N, 2$\times$600s)
observations of the giant HII region NGC 3603.
The Planetary Camera (PC) chip was centered on the bipolar outflow structure 
around the blue supergiant Sher 25. The three Wide Field Camera (WF)
chips covered the central cluster and the HII region to the south of Sher 25.

In addition, we retrieved and analyzed archival HST data, which had 
originally been obtained in July 1997 (PI Drissen). The PC was centered 
on the cluster, and the three WF chips covered the area north-west 
of the cluster. Using IRAF\footnote{IRAF is distributed
by the National Optical Astronomy Observatories (NOAO)}, we combined
individual short exposure in F547M (8$\times$30s), F675W (8$\times$20s),
and F814W (8$\times$20s) to produce images with effective exposure
times of 240s, 160s, and 160s, respectively.

The surface brightness of the proplyds was measured using aperture
photometry with an aperture radius of 0.5$''$. The photometric
calibration was carried out following the steps outlined in the
HST Data Handbook Version 3. No attempt was made to correct for
the contribution of the [\NII] lines to the H$\alpha$ F656N filter, or the
contribution of the H$\alpha$ line to the [\NII] F658N filter.
The spectrum of Proplyd 1 (see below) indicates that the underlying
continuum emission from the proplyd is negligible. Applying
equation (3) from O'Dell (1998) yields that the contribution of the 
[\NII] lines to the total flux observed in the H$\alpha$ F656N filter is 
at most 3.5\%.

\subsection{Preparatory ground-based observations}

A first set of deep ground-based broad and narrow-band images of
NGC 3603 was obtained on April 22, 1991 with the ESO New Technology
Telescope and the ESO MultiMode Instrument (EMMI). These data
were used to identify a number of compact emission nebula
in the vicinity of the central cluster.

On April 2, 1994 we tried to resolve the inner structure 
of the compact nebulae using the ESO Adaptive Optics system ADONIS.
This attempt failed due to the lack of sufficiently bright stars suitable
for wavefront sensing within 20$''$ of any of the proplyds.

A low-dispersion spectrum of Proplyd 1 was obtained on 
February 3, 1997 with the ESO/MPI 2.2m telescope and the
ESO Faint Object Spectrograph 2 (EFOSC2) at La Silla, Chile. 
The slit width was 1.5$''$. The spectrum has a sampling of 0.2 nm 
pixel$^{-1}$, a spectral resolution around 450 km\,s$^{-1}$,
and covers the wavelength range from 517 nm to 928 nm.
It was wavelength and flux calibrated using IRAF. We did not try
to correct for fringes, which become noticeable redward of 750\,nm.

\subsection{VLT/ISAAC observations}

As part of a study of the low-mass stellar content of the starburst cluster 
(see Brandl et al.\ 1999),
NGC 3603 was observed with the ESO Very Large Telescope
(VLT) Unit Telescope 1 (UT1, now officially named ``ANTU'') during the nights
of April 4--6 and 9, 1999. The observations were
carried out in service mode and used the Infrared Spectrograph And
Array Camera (ISAAC, see Moorwood et al.\ 1998).
Deep near-infrared observations of 
NGC 3603 were obtained with effective exposure times in
J$_s$, H, and K$_s$ of 2,230s, 2,710s, and 2,890s, respectively.
The seeing (FWHM) on the co-added frames was of the order of 0.35$''$
to 0.40$''$. Dithering between individual exposures increased the
field of view from its nominal value of 2.5$' \times$2.5$'$ to
3.5$' \times$3.5$'$.
The data were flux calibrated based on observations of
faint near-infrared standard stars from the lists by Hunt et al.\ (1998) and
Persson et al.\ (1998).

More details on the data reduction and analysis can be found in
Brandl et al.\ (1999).

\section{Physical properties of Proplyds}

\subsection{Morphology \& Size}

The HST/WFPC2 observations are presented in Figure \ref{fig1}.
The figure shows an overlay of two composite color images. The
upper part of the image consists of the archive data with the following
color coding: F547M (blue), F675W (green), F814W (red). Overlaid
are our new WFPC2 data with the F656N data in the red channel,
the average of F656N and F658N in the green channel, and F658N
in the blue channel.
The locations of the proplyds are marked by small boxes, and enlargements
of the boxes are shown in the upper part of Figure \ref{fig1}.
Proplyd 3 has only been observed in intermediate and broad-band
filters and thus stands out less clearly against the underlying background
when compared to Proplyds 1 and 2.
The insert at the lower right shows a color composite of HST/WFPC2
F656N (blue) and F658N (green) data and VLT/ISAAC K$_{\rm s}$ data
(red).

All three proplyds are tadpole shaped and rim brightened, with the extended
tails facing away from the starburst cluster. The portion of the ionized
rims pointing towards the cluster are brighter than the rims on the opposite
side. The central parts of the proplyds are fainter than the rims, with a
noticeable drop in surface brightness between the head and the tail.

Proplyds 2 and 3 exhibit a largely axisymmetric morphology,
whereas Proplyd 1, which is also the one closest to the cluster,
has a more complex structure. Unlike the convex shape of 
the heads of the other proplyds, Proplyd 1 has a heart-shaped head
with a collimated, outflow-like structure in between.
One possible explanation for the more complex morphology of Proplyd 1
might be that it is actually a superposition of two (or maybe even three) 
individual proplyds or that the photoevaporative flows of several disks
in a multiple system interact to produce this complex single structure.

At distances of 7.4$''$ and 2.9$''$ from Proplyd 1 and 2, respectively,
faint arc-like H$\alpha$ emission features are seen on the WFPC2
frames. The arcs are located in the direction of the cluster, and
may be the signatures of bow shocks created
by the interaction of proplyd winds with the winds from the
massive stars in the central cluster.

The proplyd heads have
diameters between 1.2$''$ and 1.7$''$ (7,200 and 10,800 A.U.). 
The head-to-tail extent of the proplyds is between 2.5$''$ and 3.5$''$
(15,000 to 21,000 A.U.). In Orion the typical diameters of
the proplyd heads vary from 45 to 355 A.U.\ (O'Dell 1998), and the 
proplyd head in M8 has a diameter of 1,080 A.U.\ (Stecklum et al.\ 1998).
Thus, the proplyds in NGC 3603 are
20 to 30 times larger than the largest proplyds in Orion, and
7 to 10 times larger than the proplyd in M8. It should
be noted that proplyds with sizes similar to those of the Orion proplyds 
would be too small 
to be resolvable at the distance of NGC 3603, where one pixel (0.1$''$)
on the wide field CCDs of WFPC2 corresponds to 600 A.U.  The PC data
with a finer pixel scale of 0.0456$''$ per pixel (270 A.U.) reveal
indeed several faint point sources which appear to be brighter
in H$\alpha$ than in [\NII]. A detailed analysis of these sources
will be subject of a later paper (Grebel et al., in prep.).

In Orion, the size of the proplyds loosely scales with
distance from the ionizing source in the sense that proplyds further 
away from $\Theta^1$\,Ori C are larger 
(McCullough et al.\ 1995; Johnstone et al.\ 
1998, O'Dell 1998).
In NGC 3603, there is no such correlation between the size of
a proplyd and its projected distance from the cluster. If, however,
the complex structure of Proplyd 1 results from multiple
photoevaporating disks as discussed above, the size estimates
based on isolated proplyds cannot be applied.
Only if Proplyd 1 can actually be decomposed into individual, isolated
proplyds with diameters around 0.9$''$, would there be a tendency
for increasing proplyd size with increasing distance from the cluster.

Coordinates, distance from the cluster center, and approximate size of
the proplyds are given in Table \ref{tab1}.

\subsection{Surface Brightness of Proplyds}

The WFPC2 observations reveal that only the outermost
layer of Proplyd 1 is ionized, whereas the interior remains neutral.
Table \ref{tab2} gives the H$\alpha$ flux and the
 surface brightness of the proplyds as measured
from the WFPC2 frames. No H$\beta$ observations were available
which would have allowed us to determine the extinction towards
the proplyds based on the Balmer decrement. Literature values
for the foreground extinction towards NGC 3603 range from
A$_{\rm v}$ = 4$^{\rm mag}$ to 5$^{\rm mag}$ 
(Moffat 1983; Melnick et al.\ 1989).
The H$\alpha$ flux has thus been corrected for
an assumed foreground extinction of A$_{\rm H\alpha}$ = 4$^{\rm mag}$. The
values for the surface brightness have not been corrected for extinction.

Proplyds in Orion get fainter with increasing distance from the ionization
source. 
Proplyd 2 is about a factor of 2.8 fainter than Proplyd 1. If the projected
separation from the cluster center is comparable to the physical distance, then
Proplyd 2 should receive a factor of $({72.5''}/{43.6''})^2$ = 2.8 less UV 
photons than Proplyd 1. The remarkably good agreement between the number of 
infalling UV photons
and the brightness of the proplyd suggests that both proplyds
are ionization bounded and receive most of the ionizing UV photons directly
from the cluster.

Proplyd 3 was only observed with intermediate and broadband filters.
Its red colors (V$-$R=3.26$^{\rm m}$, R$-$I=0.67$^{\rm m}$) are
caused by a combination of the absence of any strong emission lines in
the passband of the F547M filter, foreground extinction, and possibly
the presence of an embedded central continuum source.

Results from the VLT near-infrared broad-band photometry can be found
in Table \ref{tab3}. The surface brightness of all three proplyds
increases from J$_{\rm s}$ to H to K$_{\rm s}$. The photometry
for Proplyd 1 and 3 has not been corrected for the contamination by the 
nearby point sources.

\subsection{Nearby Point Sources}

The deep near-infrared observations with the VLT reveal a
large number of faint, red point sources.
Two point sources are detected close to the head of
Proplyd 1. As can be seen in the lower right insert of Figure \ref{fig1},
one of the point sources coincides with the
location of the ionization front and may be physically associated
with Proplyd 1. Its red near-infrared colors (see Table \ref{tab3})
indicate that the source is highly embedded and/or has a strong intrinsic
IR excess. This is in agreement with what one would expect for
a young stellar object (YSO) surrounded by a circumstellar disk and an
infalling envelope. Comparing with theoretical pre-main-sequence
 evolutionary tracks by Palla \& Stahler (1993), we note that a 1\,Myr old YSO
with J=15.4\,mag and located in NGC 3603 should have a mass around
3\,M$_\odot$ (Eisenhauer et al.\ 1998; Brandl et al.\ 1999). The effects
of additional local extinction and accretion luminosity would of course
alter this value.
In any case it should be kept in mind that the relatively high density of 
field sources makes the physical association between this point source and 
Proplyd 1 somewhat uncertain.

The point source close to the head of
Proplyd 3 is already detected on the broad-band HST/WFPC2 observations
(see Figure \ref{fig1}). The WFPC2 images show that the point source
is actually located in front of Proplyd 3 and thus very likely not
physically associated with it.

The J$_{\rm s}$, H, and K$_{\rm s}$ magnitudes of the point sources
associated with Proplyd 1 and 3 can also be found in Table \ref{tab3}.
No central point source is detected in Proplyd 2 with a limiting
magnitude of K$_{\rm s}\le$18.0\,mag.

\begin{figure*}[ht]
\psfig{figure=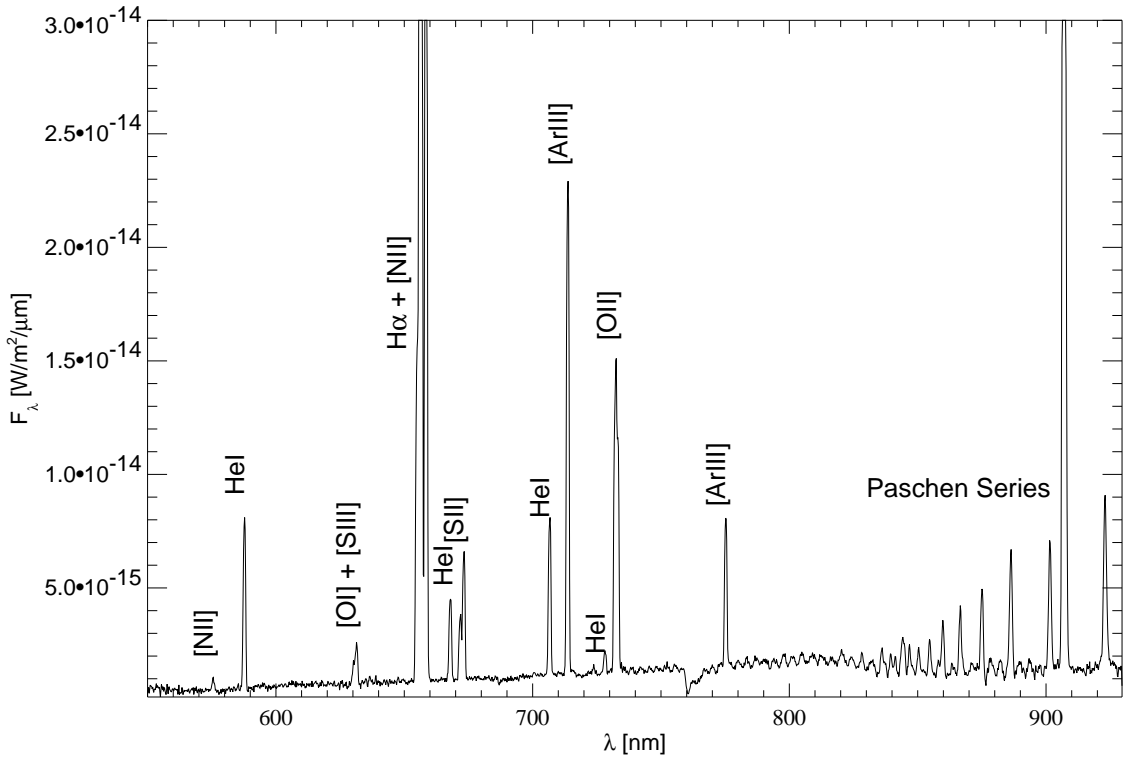,width=15cm,angle=90}
\caption{Spectrum of Proplyd 1. The most prominent emissison
lines have been marked. The rising continuum towards longer wavelengths
indicates the presence of an embedded continuum source.\label{fig2}  }
\end{figure*}

\subsection{Optical Spectroscopy}

Figure 2 shows the low-resolution spectrum of Proplyd 1. The
most prominent emission lines are identified. If one assumes
an electron temperature of 10$^4$\,K, which is quite typical for HII regions,
the flux ratio between the [S{\sc ii}] lines at 671.7\,nm and at 673.1\,nm
yields an electron density well in excess of 10$^4$ cm$^{-3}$.
Unfortunately, such densities are close to the collisional de-excitation 
limit of the [S{\sc ii}] doublet, which prevents us form getting a more
precise estimate. Other density indicators such as the extinction
corrected H$\alpha$ surface brightness or the ratio of the [C{\sc iii}]
UV doublet would have to be employed in order to derive a more accurate
estimate of the density (see Henney \& O'Dell 1999).
Densities of the order of 10$^5$ cm$^{-3}$ to
10$^6$ cm$^{-3}$ have been found in the Orion proplyds 
(Henney \& O'Dell 1999), and are also typical for Ultra Compact HII regions.

The infrared source in Proplyd 1 is also detected as 
an underlying, heavily reddened continuum source in the spectrum.
A more detailed analysis of ground based optical images and
spectra of the proplyds and other compact nebulae in NGC 3603 
will be presented in Dottori et al.\ (2000).

\subsection{Limit on the radio brightness}

The proplyds in Orion have been detected with the Very Large Array
as radio continuum sources at 2\,cm and 20\,cm (Churchwell et al.\ 1987;
Felli et al.\ 1993; McCullough et al.\ 1995).

NGC 3603 has recently been studied with the Australia Telescope Compact Array
(ATCA) at 3.4\,cm radio continuum and at several recombination emission 
lines (De Pree et al.\ 1999). Down to the 5 $\sigma$ level of 55 mJy 
beam$^{-1}$ for a beamsize of 7$''$ none of the proplyds are detected in 
the continuum.

Following the derivation by McCullough et al.\ (1995), Section 5,
the expected ratio between the radio flux density due to Bremsstrahlung,
and the H$\alpha$ flux is
$$\frac{\rm F_{\rm radio}}{\rm I_{\rm H\alpha}} = 
3.46 \nu^{-0.1}_{\rm GHz} {\rm T^{0.55}_4}
\frac{\rm mJy}{\rm photons\ cm^{-2}\ s^{-1}}~, $$
where $\nu_{\rm GHz}$ is the radio frequency in GHz and T$_4$ is the
electron temperature of the ionized gas divided by 10,000\,K.

Thus, for the H$\alpha$ flux values given in Table \ref{tab2} and T$_4$=1,
one would expect 3.4\,cm (8.8\,GHz) radio flux densities of 1.6\,mJy and 
0.55\,mJy for proplyds 1 and 2, respectively. This is well below the
detection limit of the ATCA observations by De Pree et al.\ (1999).
Internal absorption of H$\alpha$ photons emitted from the far side
of the proplyd amounts to on average less than 25\% to the total
flux (McCullough 1993) and has thus been neglected in the above estimate.

These analytical estimates are in good quantitative agreement with the
results from the numerical simulations described in Section 5.

\section{Environment}

\subsection{UV radiation field}

NGC 3603, containing over 20 O stars and WR stars, creates a much
more extreme UV environment than the Trapezium system. The
central cluster in NGC 3603 has a Lyman continuum flux of
10$^{51}$\,s$^{-1}$ (Kennicutt 1984; Drissen et al.\ 1995),
about 100 times the ionizing power of the Trapezium system.

At the same time, however, the proplyds in NGC 3603 are
at larger distances from the ionizing source. The projected separations
between the proplyds and the cluster center range from 1.3\,pc to 2.2\,pc.
The separation between $\Theta^1$\,Ori C and the Orion proplyds studied
by O'Dell (1998) varies between 0.01\,pc and 0.15\,pc.
On average, the proplyds in NGC 3603 are exposed to a somewhat less
intense EUV (h$\nu$$\ge$ 13\,eV) radiation field than the proplyds in Orion.

The most massive stars in NGC 3603 are O3V stars and WR stars
(Moffat et al.\ 1994; Drissen et al.\ 1995), which are
of earlier spectral type than the late O stars in the Trapezium system.
As a consequence, the spectral characteristics of the UV field in
NGC 3603 are different from those of the UV field in Orion. The early O-type
stars in NGC 3603,  although more luminous than the late O-type
stars in the Trapezium system, produce a disproportionate smaller rate
of FUV (13\,eV $>$ h$\nu$ $\ge$ 6\,eV) photons compared to the rate of 
EUV photons. Table \ref{tab4} gives the ratio of the FUV to EUV
photon rates for blackbodies with temperatures between 30,000\,K
and 45,000\,K.
Model calculations of the atmospheres of hot stars made available in 
electronic form by Adalbert Pauldrach\footnote{http://www.usm.uni-muenchen.de/people/adi/adi.html}
(see also Pauldrach et al.\ 1998) indicate that for effective temperatures 
below 45,000\,K the ratio of FUV to EUV photon rates  is considerably higher
than the ratio derived for a blackbody.
The models indicate ratios of 6.9:1 and 2.4:1 for O dwarfs with solar 
metallicity for effective temperatures of 35,000\,K and 40,000\,K,
respectively (see Table \ref{tab4}).

Because all the EUV photons get absorbed in the ionization front,
the FUV photon rate determines the heating inside the proplyd
envelope and thus ultimately the mass-loss rate (Johnstone et al.\ 1998).

\subsection{Winds and mass-loss rates}

Another important constituent of the environment of massive stars
are fast stellar winds and wind-wind interactions.
 In Orion, four of the five proplyds closest to $\Theta^1$\,Ori C
show arc-like features which may be bow shocks resulting from the 
interaction of the evaporation flow of the proplyds with the fast stellar 
wind from $\Theta^1$\,Ori C (McCullough et al.\ 1995).
The WFPC2 observations in H$\alpha$ of
NGC 3603 reveal similar arc-like features in front of Proplyd 1 and 2,
that might also be bow shocks.

For a stationary shock, pressure equilibrium exists on both sides of the 
shock:
$$P = \rho_{\rm cl}(r_{\rm cl}) v_{\rm cl}^2 =
     \rho_{\rm pr}(r_{\rm pr}) v_{\rm pr}^2~,$$ 
where $\rho(r)$
is the density of the wind at a distance $r$ from the cluster (cl) or
the proplyd (pr), respectively, and $v$ is the terminal velocity of the
freely expanding wind.

If radiative cooling is not important, the stagnation
point of the shock along the line connecting
the cluster with the proplyd relates to $\dot{M} v$ like

$$\frac{{\dot{M}}_{\rm cl} v_{\rm cl}}{{\dot{M}}_{\rm pr} v_{\rm pr}}
= \frac{r_{\rm cl}^2}{r_{\rm pr}^2}~,$$
where $\dot{M}$ is the mass-loss rate of the cluster and
the proplyd, respectively (e.g., Kallrath 1991).

Mass-loss rates of individual O stars in NGC 3603 are of the order of
a few 10$^{-7}$M$_\odot$\,yr$^{-1}$, and wind velocities are of the order
of a few 1,000 km\,s$^{-1}$ (Pauldrach et al.\ 1998). If we assume that the 20 O
and WR stars in NGC 3603 have  average mass-loss rates
of 3$\times$10$^{-7}$M$_\odot$\,yr$^{-1}$, and average wind velocities of 
v=2,000\,km\,s$^{-1}$, the resulting rate of momentum input $\dot M v$ from
the combined winds would be 0.012\,M$_\odot$\,yr$^{-1}$ km\,s$^{-1}$.

The separation between the arc in front of Proplyd 1 and the cluster center is 
36.2$''$, and
the separation between the arc and the proplyd is 7.4$''$.
For a bow shock, the ratio between the product of mass-loss rate times
the velocity of the wind from the cluster and Proplyd 1 would thus be 24:1.

For a wind velocity of the evaporation flow from Proplyd 1
of, e.g., 25\,km\,s$^{-1}$, the  mass-loss rate would then have to be
2$\times$10$^{-5}$M$_\odot$\,yr$^{-1}$. 
Whereas this is somewhat on the high end side of the parameter
space (see section on radiation hydrodynamical simulations below),
the arc-like feature between Proplyd 1 and the cluster could indeed
be the result of a bow shock. It should be kept in mind, however, that
the mass-loss rates for the WR stars in NGC 3603 are highly uncertain,
and might be considerably higher than our present estimate.

For Proplyd 2 the separation between the arc and the proplyd is
2.9$''$, and the separation between the arc and the cluster center is
69.6$''$. Hence the ratio between the product of mass-loss rate times
the velocity of the wind from the cluster and Proplyd 2 is 580:1, which
requires much less extreme conditions. For example,
a mass-loss rate of 2$\times$10$^{-6}$M$_\odot$\,yr$^{-1}$ and a wind velocity
of 10\,km\,s$^{-1}$ would balance the wind force from the cluster at a distance
of 2.9$''$ from Proplyd 2. Such values are in the range of mass-loss rates
and flow velocities observed for the proplyds in Orion (e.g., Henney \& O'Dell
1999; Bally et al.\ 1998b).

\begin{figure*}[ht]
\hbox{
\psfig{figure=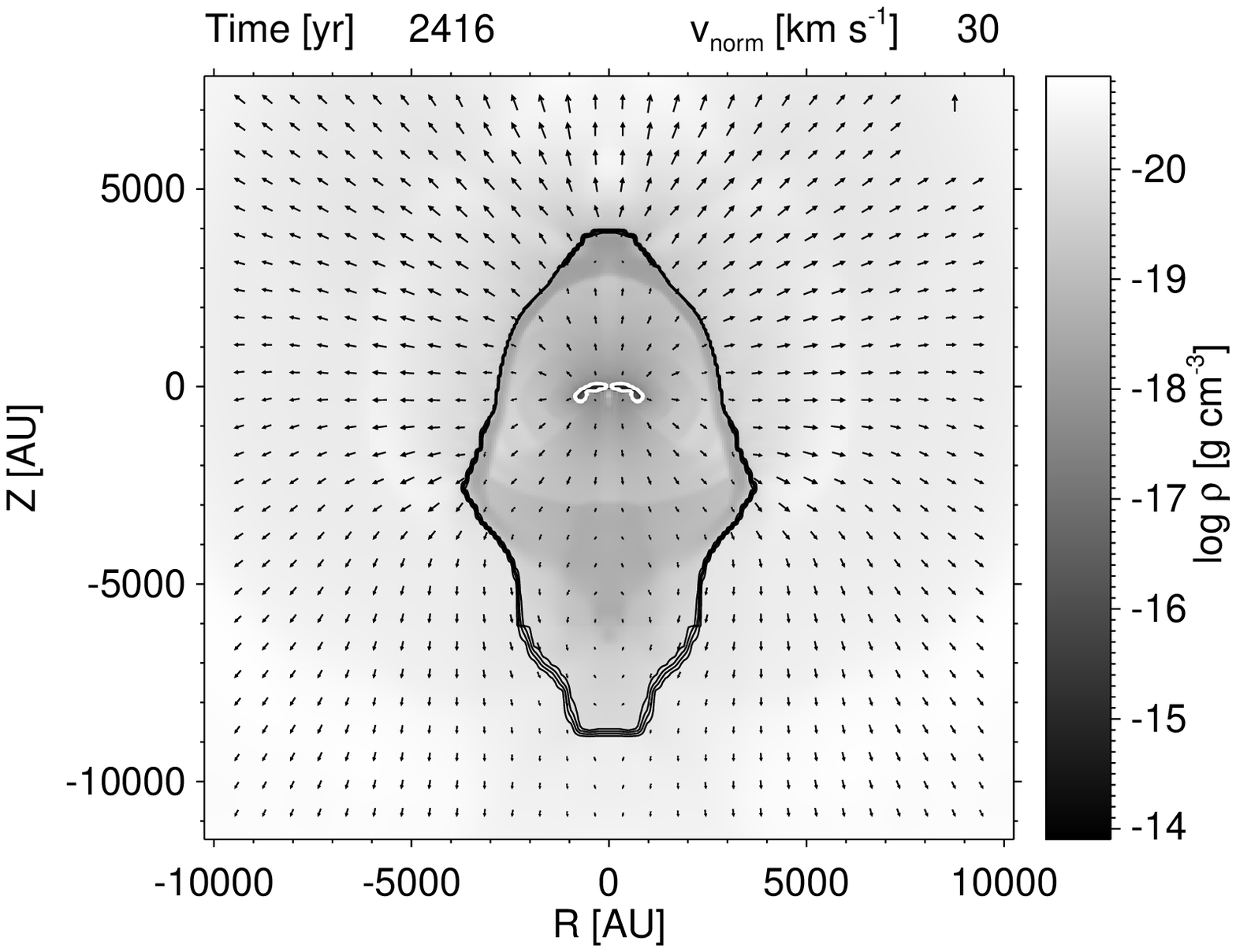,width=6cm,angle=0}
\psfig{figure=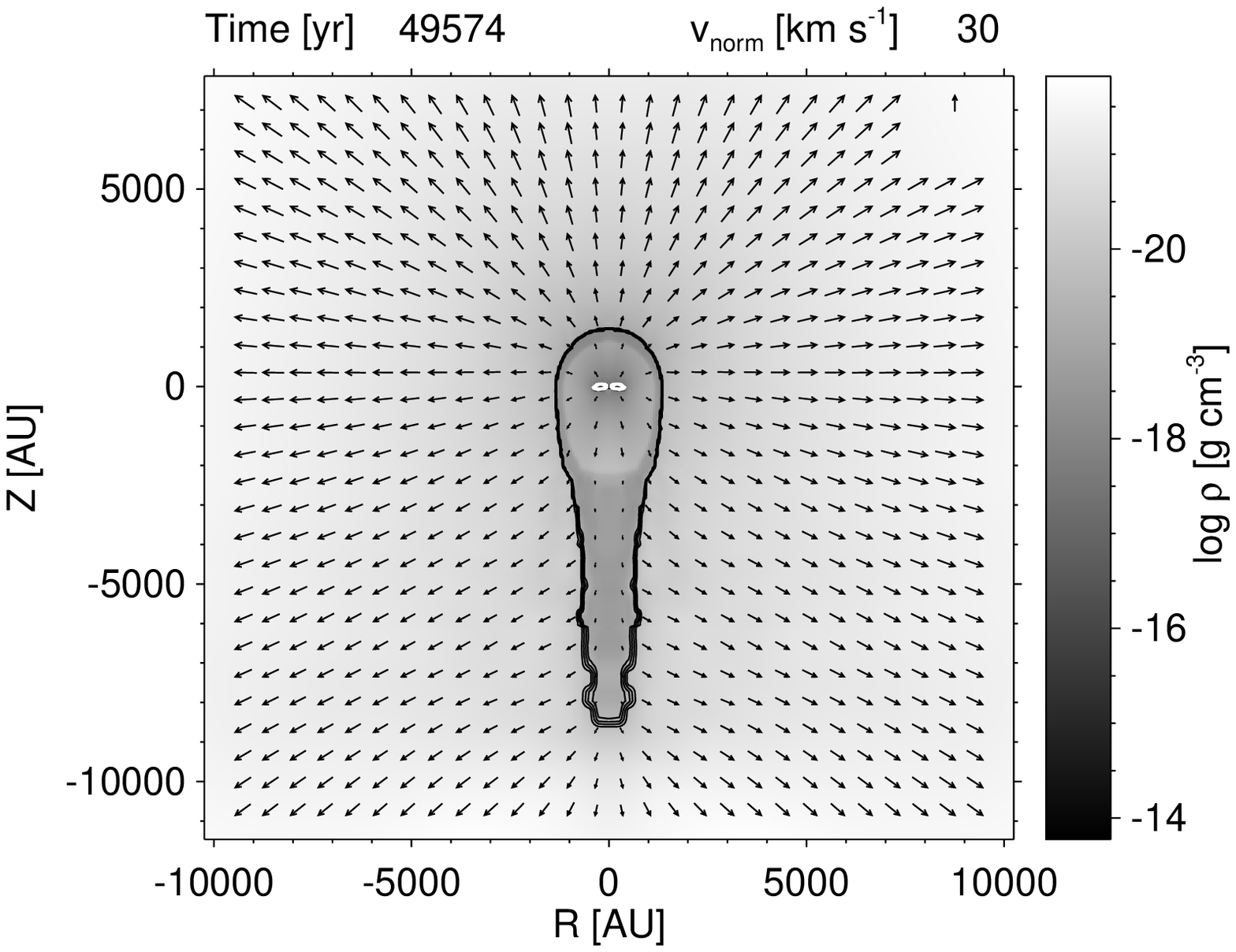,width=6cm,angle=0}
\psfig{figure=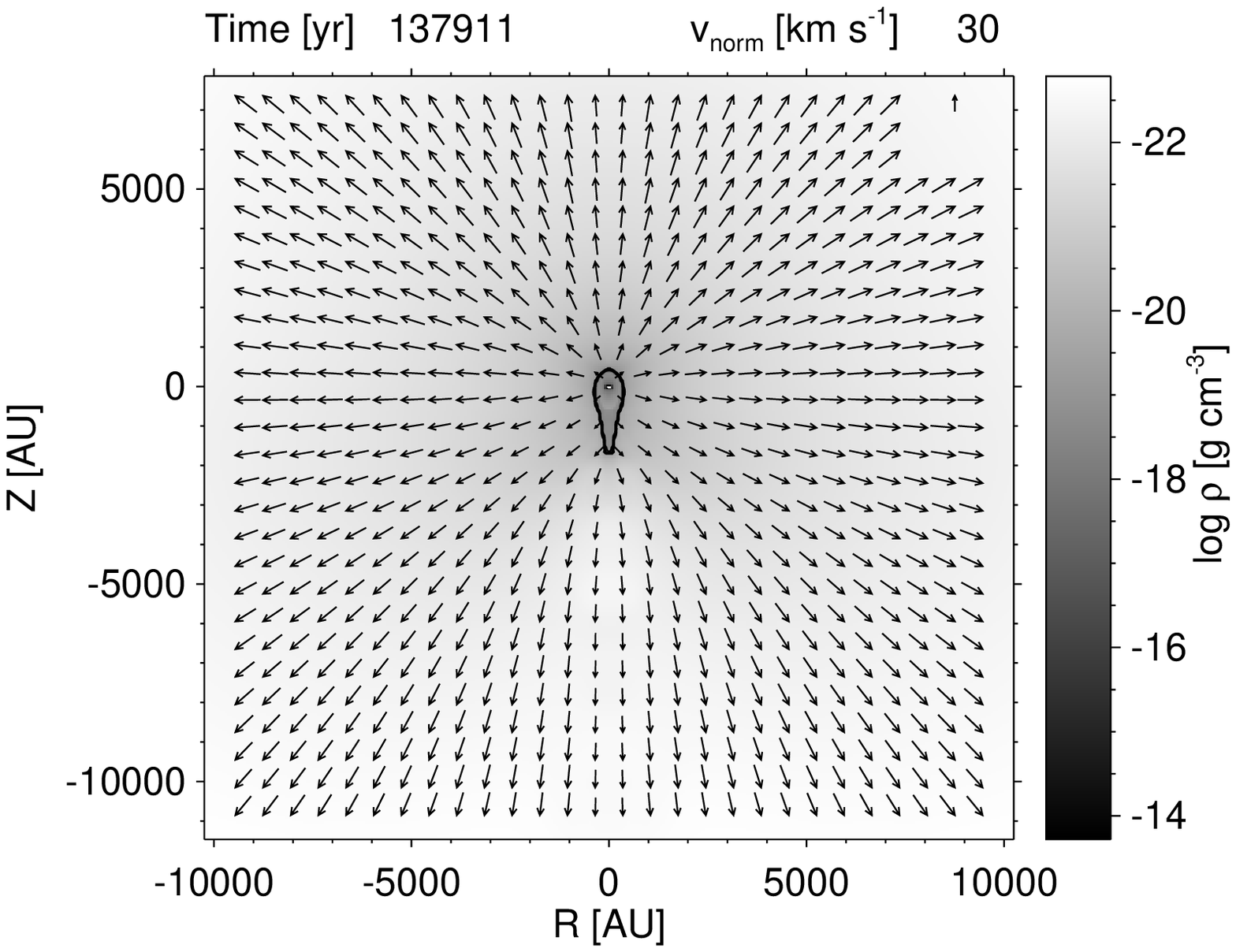,width=6cm,angle=0}
}
\caption{Time evolution of a proplyd based on
 2D radiation hydrodynamic simulations. The plots show the density
distribution and evaporation flow velocity field at t=2,416\,yr,
t$\approx$50,000\,yr, and t$\approx$138,000\,yr. The disk can be
seen as the light structure (high density) in the center of each plot.
The ionization front is indicated by the dark solid line.
 Despite an
initial mass of 0.8\,M$_\odot$ in the disk and infalling envelope, the
evaporation flow with a
mass-loss rate of 10$^{-5}$\,M$_\odot$\,yr$^{-1}$ results in
an almost complete evaporation of the disk within
$\approx$\,10$^5$\,yr.\label{fig3}}
\end{figure*}

\section{Radiation Hydrodynamical Simulations}

The proplyds in NGC 3603 are much larger than the proplyds in
Orion. In order to investigate if the proplyds in NGC 3603 can be
explained by a similar physical mechanism as the proplyds in 
Orion, we carried out radiation hydrodynamical
simulations. 

\subsection{Numerical method and initial conditions}

The simulations are based on a 2D radiation hydrodynamics code
(Yorke \& Welz 1996; Richling \& Yorke 1997, 1998; Richling 1998)
and include a thin disk with a finite scale height.
Diffuse EUV and FUV radiation fields are treated in the flux-limited diffusion
approximation (Levermore \& Pamraning 1981) as implemented for multiple
nested grids by Yorke \& Kaisig (1995). The present
simulation utilizes  6 nested grids with a resolution from 7.3 A.U.\ to
233 A.U. The coarsest outermost grid covered a cylindrical volume of
radius 13,500 A.U.\ and height 27,000 A.U.

The radial and vertical density structure of the disk was derived
from the collapse of a rotating molecular cloud core with a mass of
2\,M$_\odot$. The collapse was followed up through  5$\times$10$^5$\,yr,
at which time 1.14\,M$_\odot$ had already been accreted by the
central protostar, while 0.86\,M$_\odot$ still remained in the
protostellar disk and the infalling envelope. Angular momentum
transport was considered via an $\alpha$-prescription (Shakura \&
Sunyaev 1973), as described by Yorke \& Bodenheimer (1999).
The inclusion of the effects of angular momentum transport is
important, as disk material in the outer region of the disk
actually gains angular momentum during the evolution, which
results in a very extended disk.

The resulting, initial disk had a diameter of 3,400\,A.U. This might seem
large compared to the typical disk sizes observed for the
proplyds in Orion. Based on their numerical models,
Johnstone et al.\ (1998) derive disk sizes between 27\,A.U.\
and 175\,A.U.\ for the proplyds in 
Orion\footnote{Note that the disk radii r$_{\rm d}$ given by 
Johnstone et al.\ (1998) are actually
in units of 10$^{14}$\,cm, not in units of 10$^{17}$\,cm as erroneously
quoted in the column heads of Table 1 and 2 in their paper.}.

For disks which do not show any sign of external illumination by
UV photons, typical
disk sizes are in the range of 200 to 1,000 A.U.
(McCaughrean \& O'Dell 1996; Padgett et al.\ 1999).
It suggests that the extreme UV radiation field in the Trapezium
cluster is evaporating the disks away at a rapid pace. 
This is in agreement with the small remnant disk masses and high 
mass-loss rates, and is supported by the evaporation time scales of
the order of 10$^4$ yr computed by Henney \& O'Dell (1999).
One would expect that larger disks
also lead to the formation of more extended proplyds.
Larger disks could, e.g., be the result of an initially higher angular
momentum in the collapsing molecular cloud core.

With the aim to simulate the physical conditions close to those
observed in NGC 3603 for Proplyd 1,
the EUV photon rate was set to 10$^{51}$ photons\,s$^{-1}$,
and the distance to the ionizing source to 4.01$\times$10$^{16}$\,m (1.3\,pc). 
The effective temperature of the central source was set to 38,500\,K
and the bolometric luminosity to 2.02$\times$10$^7$\,L$_\odot$.

We carried out two sets of simulations for different FUV photon rates.
For the first set the FUV photon rate was determined from a blackbody
spectrum. As indicated in Table \ref{tab4}, this leads to an underestimate
of the true FUV photon rate and results in lower mass-loss rates and
a smaller size of the proplyd.  For the second set of simulations the
FUV photon rate was set to 1.2$\times$10$^{52}$ photons\,s$^{-1}$.
In the following we will discuss only the results from the second set
of simulations.
The initial conditions for this set of simulations are summarized in 
Table \ref{tab5}.

\subsection{Mass-loss rates and life expectancy}

Figure \ref{fig3} shows the evolution of the proplyd with time
from the instant the Lyman continuum flux was turned on.
An ionization front engulfing the star-disk-envelope system develops
almost instantaneously. Initial
mass-loss rates of the order of 10$^{-5}$\,M$_\odot$ and evaporation flow
velocities of the order of 20\,km\,s$^{-1}$ lead to a rapid depletion of the
central mass reservoir. After 50,000\,yr the disk-envelope system
has already lost 2/3 of its inital mass. After 100,000\,yr only
0.1\,M$_{\odot}$ remains in the disk. After 140,000\,yr almost 95\%
of the disk mass has been evaporated. 

The simulations confirm the finding by Johnstone et al.\ (1998)
that the FUV photon rate drives the mass-loss by heating up
the region between the embedded disk and the ionization front.
The resulting neutral flow influences the size of the ionized envelope.
The simulations agree well with time scale estimates
for disk photoevaporation by Hollenbach, Yorke, \& Johnstone (1999).

\begin{figure*}[ht]
\psfig{figure=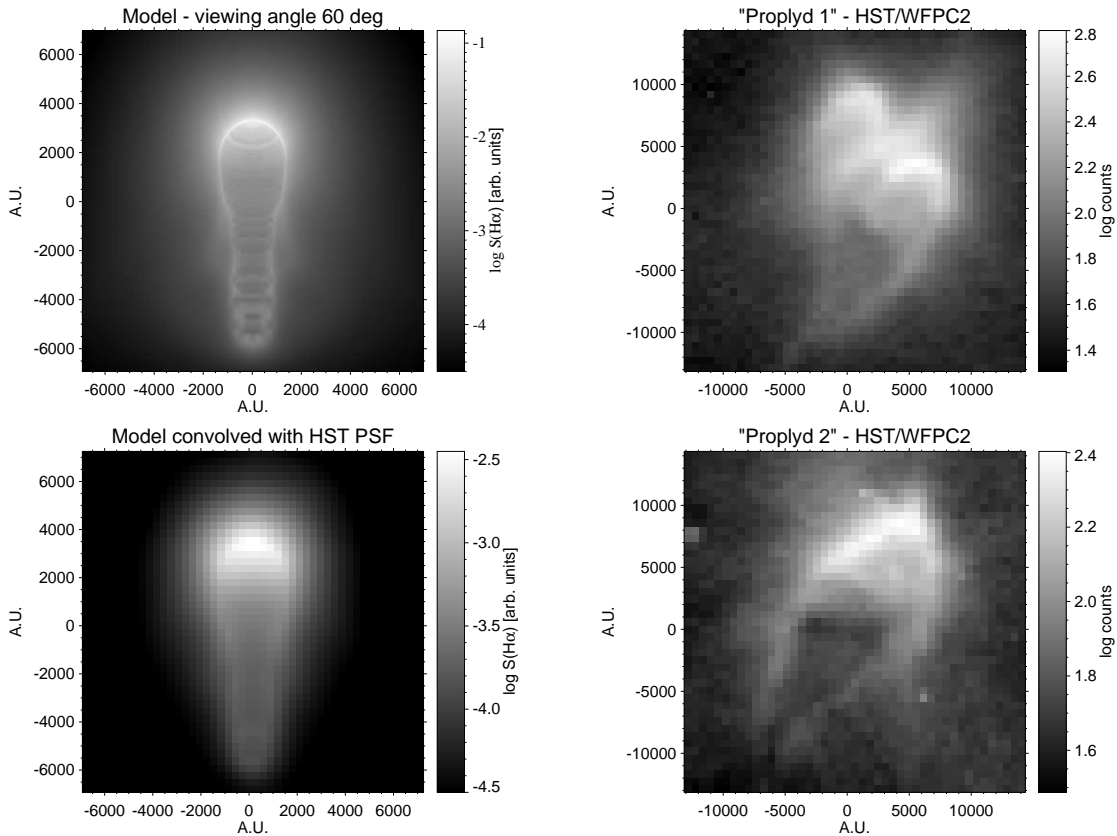,width=16cm,angle=90}
\caption{Comparison of an H$\alpha$ emission line map of
the 50,000\,yr old model and observed proplyds. The model has been convolved
with a TinyTim Point Spread Function (Krist \& Hook 1997) and resampled to the
pixelscale of the Planetary Camera.\label{fig4}}
\end{figure*}

\subsection{Emission-line maps in H$\alpha$}

In an attempt to compare the model to our HST observations, we computed 
the emission line maps in H$\alpha$. The map shown in Figure \ref{fig4}
is based on the numerical model at a time of 50,000\,yr. A viewing angle
of 60$^\circ$ was assumed.  The map was convolved with
a theoretical HST/WFPC2 point spread function computed with 
TinyTim (Krist \& Hook 1997) and then resampled to the
resolution of the Planetary Camera.

Overall, the simulated H$\alpha$ map shows a good resemblance to the 
observations, and the models provide a viable explanation for the proplyds
in NGC 3603.

Despite the high FUV photon rate, the resulting proplyd is still
a factor of two smaller than the proplyd-like structures observed in
NGC 3603. This would indicate that either the central disks are larger,
or that the proplyds in NGC 3603 have only very recently (within the 
last 10$^4$\,yr) been exposed to the UV radiation field. The latter would cause
timescale problems, as it would require a very rapidly receeding
HI/HII front.

\section{Summary and Outlook}

While the present analysis cannot give a definite answer about the
nature of the proplyd-like nebulae in NGC 3603, the WFPC2 data
already provide a number of important clues:

i) The tadpole shape suggests that the dynamics of the
proplyds are dominated by the central cluster of NGC 3603.

ii) The variation of the H$\alpha$ brightness of the proplyds with
distance to the ionizing cluster further indicates that the proplyds
are ionization bounded.

iii) The HST/WFPC2 and VLT/ISAAC images nicely resolve the nebulae
into a region of neutral material surrounded by an outer ionization front.
As discussed by O'Dell (1998), the fact that proplyds are ionization
bounded implies that no EUV photons, but only FUV photons 
penetrate and heat up the central region of the proplyds.
The neutral interior provides a steady supply of material 
for the ionization front.

iv) The key to a better understanding of the internal density structure of the
nebulae is the direct detection of central point sources, such as the
source possibly associated with Proplyd 1.  The near-infrared photometry
alone, however, does not really allow for a good estimate of the
overall luminosity of the
embedded source and the amount of internal extintion.
Similarly, the limited spectral resolution of our ground-based
spectra does not allow us to
determine mass-loss rates for the proplyds.

v) Faint nebular arcs detected between the proplyds and the cluster
can be explained by bow shocks resulting from the interaction of
the evaporation flow with the winds from the high-mass stars in the cluster.

The simulations indicate that high mass-loss rates of the order of 
10$^{-5}$\,M$_\odot$\,yr$^{-1}$ are necessary in order to explain
the physical size of the proplyds. Similar to the Trapezium system,
the proplyd-like structures in NGC 3603 appear to be short-lived
phenomena. The fact that we detect three proplyds in NGC 3603 then
suggests that proplyds/circumstellar disks are common not only in Orion,
but throughout the Milky Way.

Another consequence of the short life expectancy of proplyds in
giant HII region is that they are not likely to survive long enough to form
planetary systems, which are expected to form on timescales
of the order of 10$^6$\,yr (Lissauer 1987). It might, however,
also be possible to form giant planets on
a considerably shorter timescale (Boss 1998).

While we do not have definite proof that the proplyd-like features
in NGC 3603 are of the same nature as the proplyds in Orion, the
similarities in morphology and physical characteristics strongly suggest
that they are related phenomena. The numerical modelling shows that
the proplyds can indeed be scaled to the size and the physical
conditions in NGC 3603.

Further observations are needed to elucidate the nature of the proplyd-like
nebulae in NGC 3603.  High-spatial resolution imaging
and spectroscopy with HST in the optical and near-infrared will
provide additional insights.
Similar to Orion, it might be possible to detect the central disks
glowing in the optical [O{\sc i}] emission line, and in the 
near-infrared molecular hydrogen lines. The physical extent, location
within the proplyd,  and inclination of the disk
would provide important constraints for the simulations.

While the radio continuum survey by De Pree et al.\ (1999) at 3.4\,cm
did not detect the proplyds, in the future it might be possible to detect and
resolve the disks with the Atacama Large Millimeter Array (ALMA), which is
currently under study for the Chilean Atacama desert.

Deep ground-based thermal infrared imaging with the new generation 
of 6 to 8\,m class telescopes in the southern hemisphere could lead to the 
detection of more heavily embedded
protostars in the center of the proplyds. The luminosity of each
protostar would provide important constraints on its mass and evolutionary
status.

\acknowledgements
This research is supported by
NASA through grant number GO-07373.01-96A from the Space Telescope Science
Institute, which is operated by the Association of Universities for Research
in Astronomy, Inc., under NASA contract NAS5-26555. Part of this research has
been carried out at the Jet Propulsion Laboratory (JPL), California
Institute of Technology, and has been supported by NASA through the
``Origins'' program. The calculations were performed on computers operated by
the ``Rechenzentrum der Universit\"at W\"urzburg'', the JPL/Caltech
Supercomputing Project, and the John von Neumann Institute for
Computing in J\"ulich.
EKG acknowledges support by NASA through grant HF-01108.01-98A from the
Space Telescope Science Institute.
YHC acknowledges the NASA grant NAG 5-3246.  HD acknowledges support by a
fellowship from the Alexander-von-Humboldt-Foundation. We would like to
thank Robert Gruendl for helpful discussions and comments, and the
anonymous referee for the fast reply and the helpful
comments and suggestions.

\begin{table}[htb]
\caption{\label{tab1} Location and size of the three proplyd-like structures}
\begin{tabular}{ccccc}
Name & RA(2000) & DEC(2000) &distance from cluster & size \\ \hline
Proplyd 1   & 11h\,15m\,13.13s&-61$^\circ$\,15$'$\,50.0$''$  & 43.6$''$ (1.3\,pc) & 1.8$''^a$ $\times$ 3.2$''$\\
Proplyd 2   & 11h\,15m\,16.59s&-61$^\circ$\,16$'$\,06.2$''$  & 72.5$''$ (2.2\,pc) & 1.4$''$ $\times$ 3.5$''$\\
Proplyd 3   & 11h\,15m\,07.73s&-61$^\circ$\,15$'$\,16.8$''$  & 68.0$''$ (2.0\,pc) & 1.2$''$ $\times$ 2.5$''$\\ \hline
\end{tabular}

a:0.9$''$, if we assume that Proplyd 1 is a superposition
of two or three individual proplyds
\end{table}

\begin{table}[htb]
\caption{\label{tab2} Surface brightness of proplyd heads
in H$\alpha$ (F656N), [\NII] (F658N), V (F547M), R (F675W), and I (F814W)
measured for a 0.5$''$ aperture radius in the HST VEGAMAG system.}
\begin{tabular}{ccccccc}
Name &I$_{\rm H\alpha}$ [cm$^{-2}$s$^{-1}$]$^a$  &
 H$\alpha$ [mag/$''^2$]  &
[\NII] [mag/$''^2$] & F547M [mag/$''^2$] &
F675W [mag/$''^2$]& F814W [mag/$''^2$]\\ \hline
Proplyd 1   & 0.56& 13.60 & 15.90 &---  &--- &---\\
Proplyd 2   & 0.20& 14.73 & 16.71 &---  &--- &---\\
Proplyd 3   & ---  & --- &---  &24.84  & 21.58 &20.91\\ \hline
\end{tabular}

a: Flux corrected for an assumed foreground extinction in
H$\alpha$ of A$_{\rm H\alpha}$=4mag
\end{table}

\begin{table}[htb]
\caption{\label{tab3} Near-infrared surface brightness of proplyd heads
and photometry of nearby point sources ``*'' (VLT/ISAAC).} 
\begin{tabular}{cccc}
Name &J$_{\rm s}$ & H  & K$_{\rm s}$ \\ \hline
Proplyd 1   & 14.8 mag/$''^2$ & 13.7 mag/$''^2$ & 12.9 mag/$''^2$ \\
Proplyd 2   & 16.2 mag/$''^2$ & 15.7 mag/$''^2$ & 14.5 mag/$''^2$ \\
Proplyd 3   & 14.0 mag/$''^2$ & 12.8 mag/$''^2$ & 12.2 mag/$''^2$ \\ \hline
Proplyd 1*  & 15.4 mag & 14.1 mag & 13.4 mag \\
Proplyd 3*  & 12.9 mag & 13.2 mag & 12.6 mag \\ \hline
\end{tabular}
\end{table}

\begin{table}[htb]
\caption{\label{tab4} Ratio of FUV to EUV photon rates} 
\begin{tabular}{cccc}
T$_{\rm eff}$ [K]     & blackbody  &
\multicolumn{2}{c}{stellar photosphere$^a$ }\\[.2ex]
 &  &Z = 0.2\,Z$_\odot$ &Z = Z$_\odot$ \\ \hline
30,000& 4.9 & 45.5 & 130\\
35,000& 3.1 & 6.4 & 6.9\\
40,000& 2.2 & 2.3 & 2.4\\
45,000& 1.6 & 1.7 & 1.4\\ \hline
\end{tabular}

a: Derived from stellar photosphere models by A.\ Pauldrach
et al.\ (1998).
\end{table}

\begin{table}[htb]
\caption{\label{tab5} Model Parameters}
\begin{tabular}{rl}
distance to ionizing source & 4.01 $\times$ 10$^{16}$\,m (1.3\,pc)\\
T$_{\rm eff}$&       38,500\,K\\
Luminosity& 2.02 $\times$ 10$^7$  L$_\odot$\\
EUV flux (h$\nu$ $\ge$ 13.6\,eV)&     10$^{51}$\,s$^{-1}$ \\
FUV flux (6\,eV $\le$ h$\nu$ $<$ 13.6\,eV) & 1.2 $\times$ 10$^{52}$\,s$^{-1}$\\
\end{tabular}
\end{table}

\end{document}